\RequirePackage[mathlines]{lineno}
\documentclass[prl,twocolumn,showpacs,amsmath,amssymb]{revtex4}
\usepackage{overpic,graphicx}
\usepackage{dcolumn}
\usepackage{bm}
\usepackage{rotating}
\usepackage{subfigure}
\usepackage{color}

\begin{document}

\title{\bf \boldmath
Study of the $D^0\to K^-\mu^+\nu_\mu$ dynamics and test of lepton flavor universality with $D^0\to K^-\ell^+\nu_\ell$ decays
}

\author{
M.~Ablikim$^{1}$, M.~N.~Achasov$^{9,d}$, S. ~Ahmed$^{14}$, M.~Albrecht$^{4}$, M.~Alekseev$^{55A,55C}$, A.~Amoroso$^{55A,55C}$, F.~F.~An$^{1}$, Q.~An$^{52,42}$, J.~Z.~Bai$^{1}$, Y.~Bai$^{41}$
% ...
\\
\vspace{0.2cm}
(BESIII Collaboration)\\
\vspace{0.2cm} {\it
$^{1}$ Institute of High Energy Physics, Beijing 100049, People's Republic of China\\
% ...
$^{4}$ Bochum Ruhr-University, D-44780 Bochum, Germany\\
% ...
$^{9}$ G.I. Budker Institute of Nuclear Physics SB RAS (BINP), Novosibirsk 630090, Russia\\
% ...
$^{14}$ Helmholtz Institute Mainz, Johann-Joachim-Becher-Weg 45, D-55099 Mainz, Germany\\
% ...
$^{41}$ Southeast University, Nanjing 211100, People's Republic of China\\
$^{42}$ State Key Laboratory of Particle Detection and Electronics, Beijing 100049, Hefei 230026, People's Republic of China\\
% ...
$^{52}$ University of Science and Technology of China, Hefei 230026, People's Republic of China\\
% ...
$^{55}$ (A)University of Turin, I-10125, Turin, Italy; (B)University of Eastern Piedmont, I-15121, Alessandria, Italy; (C)INFN, I-10125, Turin, Italy\\
% ...
$^{d}$ Also at the Novosibirsk State University, Novosibirsk, 630090, Russia\\
% ...
}
}

\begin{abstract}
  Using $e^+e^-$ annihilation data of $2.93~\mathrm{fb}^{-1}$
  collected at center-of-mass energy $\sqrt{s}=3.773$ GeV with the
  BESIII detector, we measure the absolute branching fraction of
  $D^{0}\to K^{-}\mu^{+}\nu_{\mu}$ with significantly improved
  precision: ${\mathcal B}_{D^{0}\to
    K^{-}\mu^{+}\nu_{\mu}}=(3.413\pm0.019_{\rm stat.}\pm0.035_{\rm
    syst.})\%$.  Combining with our previous measurement of ${\mathcal
    B}_{D^0\to K^-e^+\nu_e}$, the ratio of the two branching fractions
  is determined to be ${\mathcal B}_{D^0\to K^-\mu^+\nu_\mu}/{\mathcal
    B}_{D^0\to K^-e^+\nu_e}=0.974\pm0.007_{\rm stat.}\pm0.012_{\rm
    syst.}$, which agrees with the theoretical expectation of lepton flavor
  universality within the uncertainty.  A study of the ratio of the
  two branching fractions in different four-momentum transfer regions
  is also performed, and no evidence for lepton flavor universality violation
  is found with current statistics.  Taking inputs from global fit in the 
  standard model and lattice quantum chromodynamics separately, we 
  determine $f_{+}^{K}(0)=0.7327\pm0.0039_{\rm
    stat.}\pm0.0030_{\rm syst.}$ and $|V_{cs}| = 0.955\pm0.005_{\rm
    stat.}\pm0.004_{\rm syst.}\pm0.024_{\rm LQCD}$.

\end{abstract}


\maketitle

\oddsidemargin  -0.2cm
\evensidemargin -0.2cm

In the standard model (SM), lepton flavor universality (LFU) requires equality of couplings between three families of leptons and gauge bosons. Semileptonic (SL) decays of pseudoscalar mesons, well understood in the SM, offer an excellent opportunity to test LFU and search for new physics effects. Recently, various LFU tests in SL $B$ decays were reported at BaBar, Belle and LHCb. The measured branching fraction (BF) ratios
${\mathcal R}_{D^{(*)}}^{\tau/\ell}={\mathcal B}_{B\to \bar D^{(*)}\tau^+\nu_\tau}/{\mathcal B}_{B\to \bar D^{(*)}\ell^+\nu_\ell}$~($\ell=\mu$, $e$)~\cite{babar_1,babar_2,lhcb_1,belle2015,belle2016}
and
${\mathcal R}_{K^{(*)}}^{\mu\mu/ee}={\mathcal B}_{B\to K^{(*)}\mu^+\mu^-}/{\mathcal B}_{B\to K^{(*)}e^+e^-}$~\cite{lhcb_kee_3,lhcb_kee_4}
deviate from SM predictions by $3.9\sigma$~\cite{HFLAV} and 2.1-$2.5\sigma$, respectively.
Various models~\cite{BFajfer2012,Fajfer2012,Celis2013,Crivellin2015,Crivellin2016,Bauer2016}
were proposed to explain these tensions. Precision measurements of SL $D$ decays provide critical and complementary tests of LFU. Reference~\cite{Fajfer2015} states that observable LFU violations may exist in $D^0\to K^-\ell^+\nu_\ell$ decays. In the SM, Ref.~\cite{Riggio2018} predicts ${\mathcal R}_{\mu/e}={\mathcal B}_{D^0\to K^-\mu^+\nu_\mu}/{\mathcal B}_{D^0\to K^-e^+\nu_e}=0.975\pm0.001$.
Above $q^2=0.1$\,GeV$^2/c^4$ ($q$ is the total four momentum of $\ell^+\nu_\ell$), one expects ${\mathcal R}_{\mu/e}$ close to 1 with negligible uncertainty~\cite{ETM2017}. This Letter presents an improved measurement of $D^0\to K^-\mu^+\nu_\mu$~\cite{charge}, and LFU test with $D^0\to K^-\ell^+\nu_\ell$ decays in the full kinematic range and various separate $q^2$ intervals.

Moreover, experimental studies of the $D^0\to K^-\ell^+\nu_\ell$ dynamics help to determine the $c\to s$ quark mixing matrix element $|V_{cs}|$ and the hadronic form factors (FFs) $f_\pm^K(0)$~\cite{Riggio2018,Zhang2018,Fang2015}. The $D^0\to K^-e^+\nu_e$ dynamics was well studied by CLEO-c, Belle, BaBar, and BESIII~\cite{belle2006,cleo2009,babar2007,bes2015}. However, the $D^0\to K^-\mu^+\nu_\mu$ dynamics was only investigated by Belle and
FOCUS~\cite{focus2005,belle2006}, with relatively poor precision. By analyzing the $D^0\to K^-\mu^+\nu_\mu$ dynamics, we determine $|V_{cs}|$ and $f^K_+(0)$ incorporating the inputs from global fit in the SM~\cite{pdg2016} and lattice quantum chromodynamics\,(LQCD)~\cite{LQCD}. These are critical to test quark mixing matrix unitarity and validate LQCD calculations on FFs.
This analysis is performed using 2.93~fb$^{-1}$ of data taken at center-of-mass energy $\sqrt s=$ 3.773~GeV with the
BESIII detector.

Details about the design and performance of the BESIII detector are
given in Ref.~\cite{BESCol}.  The Monte Carlo~(MC) simulated events are
generated with a {\sc geant4}-based~\cite{geant4} detector simulation
software package, {\sc boost}. An inclusive MC sample, which includes
the $D^0\bar{D}^0$, $D^+D^-$ and non-$D\bar{D}$ decays of
$\psi(3770)$, the initial state radiation~(ISR) production of
$\psi(3686)$ and $J/\psi$, and the $q\bar{q}$~($q=u,d,s$) continuum
process, along with Bhabha scattering, $\mu^+\mu^-$ and $\tau^+\tau^-$
events, is produced at $\sqrt{s}=3.773$~GeV to determine the detection
efficiencies and to estimate the potential backgrounds. The production
of the charmonium states is simulated by the MC generator {\sc
kkmc}~\cite{kkmc}.  The measured decay modes of the charmonium
states are generated using {\sc evtgen}~\cite{evtgen} with BFs from
the Particle Data Group~(PDG)~\cite{pdg2016}, and the remaining
unknown decay modes are generated by {\sc lundcharm}~\cite{lundcharm}.
The $D^0\to K^-\mu^+\nu_\mu$ decay is simulated with the modified pole
model~\cite{MPM}.

At $\sqrt{s}=3.773$~GeV, the $\psi(3770)$ resonance decays
predominately into $D^0\bar{D}^0$ or $D^+D^-$ meson pairs. If a
$\bar{D}^0$ meson is fully reconstructed by $\bar{D}^0\to K^+\pi^-,
K^+\pi^-\pi^0$ and $K^+\pi^-\pi^-\pi^+$, a $D^0$ meson must exist in
the recoiling system of the reconstructed $\bar{D}^0$~(called the
single-tag (ST) $\bar{D}^0$).  In the presence of the ST $\bar D^0$,
we select and study $D^0\to K^-\mu^+\nu_\mu$ decay~(called the
double-tag (DT) events).  The BF of the SL decay is given by
\begin{equation}
\label{eq:bf}
{\mathcal B}_{D^{0}\to K^-\mu^+\nu_\mu}=N_{\mathrm{DT}}/(N_{\mathrm{ST}}^{\rm tot}\times\varepsilon_{\rm SL}),
\end{equation}
where $N_{\rm ST}^{\rm tot}$ and $N_{\rm DT}$ are the ST and DT
yields, $\varepsilon_{\rm SL}=\varepsilon_{\rm DT}/\varepsilon_{\rm
  ST}$ is the efficiency of reconstructing $D^0\to K^-\mu^+\nu_\mu$ in
the presence of the ST $\bar{D}^{0}$, and $\varepsilon_{\rm ST}$ and
$\varepsilon_{\rm DT}$ are the efficiencies of selecting ST and DT
events.

All charged tracks must originate from the interaction point with a
distance of closest approach less than 1~cm in the transverse plane
and less than 10~cm along the $z$ axis. Their polar angles ($\theta$)
are required to satisfy $|\cos\theta|<0.93$.  Charged particle
identification~(PID) is performed by combining the time-of-flight
information and the specific ionization energy loss measured in the
main drift chamber. The information of the electromagnetic
calorimeter~(EMC) is also included to identify muon
candidates. Combined confidence levels for electron, muon, pion and
kaon hypotheses~($CL_e$, $CL_\mu$, $CL_\pi$ and $CL_K$) are calculated
individually.  Kaon (pion) and muon candidates must satisfy
$CL_{K(\pi)}>CL_{\pi(K)}$ and $CL_{\mu}>$ 0.001, $CL_e$ and $CL_K$,
respectively.  In addition, the deposited energy in the EMC of the muon
is required to be within (0.02,\,0.29)\,GeV.  The $\pi^0$ meson is
reconstructed via $\pi^0\to\gamma\gamma$ decay.  The energy deposited
in the EMC of each photon is required to be greater than 0.025~GeV in
the barrel ($|\cos\theta|<0.80$) region or 0.050~GeV in the end cap
($0.86<|\cos\theta|<0.92$) region, and the shower time has to be
within 700 ns of the event start time.  The $\pi^0$ candidates with
both photons from the end cap are rejected because of poor
resolution. The $\gamma\gamma$ combination with an invariant mass
($M_{\gamma\gamma}$) in the range $(0.115,\,0.150)$\,GeV$/c^{2}$ is
regarded as a $\pi^0$ candidate, and a kinematic fit by constraining
the $M_{\gamma\gamma}$ to the $\pi^0$ nominal mass~\cite{pdg2016} is
performed to improve the mass resolution.  For $\bar{D}^0\to
K^+\pi^-$, the backgrounds from cosmic ray events, radiative Bhabha
scattering and dimuon events are suppressed with the same
requirements as used in Ref.~\cite{cosmic}. 

\begin{figure}[htbp]\centering
\includegraphics[width=0.45\textwidth]{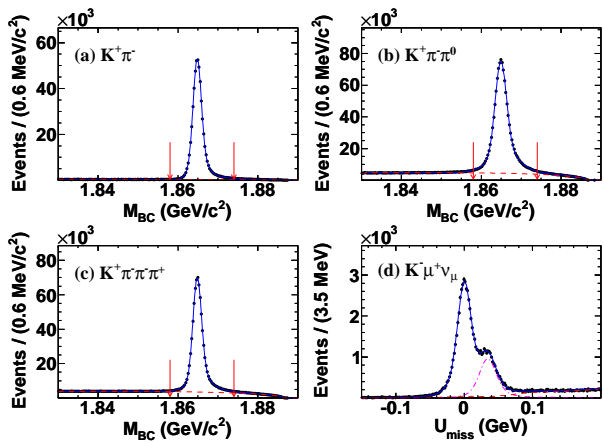}
\caption{Fits to (a-c) the $M_{\rm BC}$ distributions for the three ST
  modes, and (d) the $U_{\rm miss}$ distribution for $D^0\to
  K^-\mu^+\nu_\mu$ candidates. Dots with error bars are data, solid
  curves show the fit results, dashed curves show the fitted
  non-peaking background shapes, the dash-dotted curve in (d) is the
  peaking background shape of $D^0\to K^-\pi^+\pi^0$ and the red
  arrows in (a-c) give the $M_{\rm BC}$ windows.}
\label{fig:fit}
\end{figure}

The ST $\bar{D}^0$ mesons are identified by the energy difference
$\Delta E\equiv E_{\bar D^0}-E_{\rm beam}$ and the
beam-constrained mass $M_{\rm
  BC}\equiv\sqrt{E_{\mathrm{beam}}^{2}-|\vec{p}_{\bar{D}^0}|^{2}}$,
where $E_{\mathrm{beam}}$ is the beam energy, and $E_{\bar D^0}$ and
$\vec{p}_{\bar{D}^0}$ are the total energy and momentum of the ST
$\bar{D}^0$ in the $e^+e^-$ rest frame. If there are multiple combinations in an
event, the combination with the smallest $|\Delta E|$ is chosen for each tag
mode and for $D^0$ and $\bar D^0$. For one event, there may be up to six ST $D$ candidates selected.
To determine the ST yield, we fit the $M_{\rm BC}$
distributions of the accepted candidates after imposing mode dependent
$\Delta E$ requirements.  The signal is described by the MC-simulated
shape convolved with a double-Gaussian function accounting for the
resolution difference between data and MC simulation, and the
background is modeled by an ARGUS function~\cite{argus}.  Fit results
are shown in Figs.~\ref{fig:fit}(a-c).
The corresponding $\Delta E$ and $M_{\rm BC}$ requirements, ST yields and efficiencies for various ST modes are summarized in Table~\ref{tab:styields}. The total ST yield is $N^{\rm tot}_{\rm ST}=2341408\pm2056$.	

\begin{table*}[htbp]\centering
\caption{$\Delta E$ and $M_{\rm BC}$ requirements, ST yields $N_{\rm ST}$, ST efficiencies $\varepsilon_{\rm ST}$ and signal efficiencies $\varepsilon_{\rm SL}$ for different ST modes. Uncertainties are statistical only.}
\label{tab:styields}
\begin{tabular}{lccccc}
\hline
{\rm ST} mode & $\Delta E$ (MeV) & $M_{\rm BC}$ (GeV/$c^2$) & $N_{\rm ST}$ & $\varepsilon_{\rm ST}$ (\%) & $\varepsilon_{\rm SL}$ (\%) \\\hline
$K^+\pi^-$ & $(-29,\,27)$ & $(1.858,\,1.874)$ & $538865\pm785$ & $65.37\pm0.09$ & $57.74\pm0.09$\\
$K^+\pi^-\pi^0$ & $(-69,\,38)$ & $(1.858,\,1.874)$ & $1080050\pm1532$ & $34.67\pm0.04$ & $61.23\pm0.09$\\
$K^+\pi^-\pi^-\pi^+$ & $(-31,\,28)$ & $(1.858,\,1.874)$ & $722493\pm1126$ & $38.20\pm0.06$ & $56.42\pm0.09$\\
\hline
\end{tabular}
\end{table*}

Candidates for $D^{0}\to K^-\mu^+\nu_\mu$ must contain two oppositely
charged tracks which are identified as a kaon and muon, respectively.
The muon must have the same charge as the kaon on the ST side. To
suppress the peaking backgrounds from $D^0\to K^-\pi^+(\pi^0)$, the
$K^-\mu^+$ invariant mass~($M_{K^-\mu^+}$) is required to be less than
1.56 GeV/$c^2$, and the maximum energy of any photon that is not used
in the ST selection~($E_{\rm extra~\gamma}^{\rm max}$) must be less
than 0.25 GeV.

The kinematic quantity $U_{\mathrm{miss}}\equiv
E_{\mathrm{miss}}-|\vec{p}_{\mathrm{miss}}|$ is calculated for each
event, where $E_{\mathrm{miss}}$ and $\vec{p}_{\mathrm{miss}}$ are the
energy and momentum of the missing particle, which can be calculated
by $E_{\mathrm{miss}}\equiv E_{\mathrm{beam}}-E_{K^{-}}-E_{\mu^{+}}$
and
$\vec{p}_{\mathrm{miss}}\equiv\vec{p}_{D^{0}}-\vec{p}_{K^{-}}-\vec{p}_{\mu^{+}}$
in the $e^+e^-$ center-of-mass frame, where $E_{K^-(\mu^+)}$ and
$\vec{p}_{K^{-}(\mu^+)}$ are the energy and momentum of the kaon
(muon) candidates. To improve the $U_{\mathrm{miss}}$ resolution, the
$D^0$ energy is constrained to the beam energy and $\vec{p}_{D^{0}}
\equiv -\hat{p}_{\bar
  D^0}\sqrt{E_{\mathrm{beam}}^{2}-m_{\bar{D}^{0}}^{2}}$, where
$\hat{p}_{\bar{D}^0}$ is the unit vector in the momentum direction of
the ST $\bar{D}^{0}$ and $m_{\bar{D}^{0}}$ is the $\bar{D}^0$ nominal
mass~\cite{pdg2016}.

The SL decay yield is obtained from an unbinned fit to the
$U_{\mathrm{miss}}$ distribution of the accepted events of data, as
shown in Fig.~\ref{fig:fit} (d). In the fit, the signal, the peaking
background of $D^{0}\to K^{-}\pi^{+}\pi^{0}$ decay and other
backgrounds are described by the corresponding MC-simulated
shapes. The former two are convolved with the same Gaussian function
to account for the resolution difference between data and MC
simulation. All parameters are left free. The fitted signal yield is
$N_{\rm DT}=47100\pm259$.

The efficiencies of finding $D^0\to K^-\mu^+\nu_\mu$ for different ST
modes are summarized in Table~\ref{tab:styields}. They are weighted by
the ST yields and give the average efficiency $\varepsilon_{\rm
  SL}=(58.93\pm0.07)\%$.  To verify the reliability of the efficiency,
typical distributions of the SL decay, $e.g.$, momenta and
$\cos\theta$ of $K^-$ and $\mu^+$, are checked and good consistency
between data and MC simulation has been found (See Fig.~1 of
Ref.~\cite{app}).

By inserting $N_{\rm DT}$, $\varepsilon_{\rm SL}$ and $N_{\rm ST}^{\rm tot}$ into Eq.~(\ref{eq:bf}), one obtains
$${\mathcal B}_{D^0\to K^-\mu^+\nu_\mu}=(3.413\pm0.019_{\rm stat.}\pm0.035_{\rm syst.})\%.$$
The systematic uncertainties in the BF measurement are described as
follows.  The uncertainty in $N_{\rm ST}^{\rm tot}$ is taken as 0.5\%
by examining the changes of the fitted yields by varying the fit
range, the signal shape and the endpoint of the ARGUS function.  The
efficiencies of muon and kaon tracking (PID) are studied with
$e^+e^-\to\gamma\mu^+\mu^-$ events and DT hadronic events,
respectively.  The uncertainties of tracking and PID efficiencies each
are assigned as 0.3\% per kaon or muon.  The differences of the
momentum and $\cos\theta$ distributions between $D^0\to
K^-\mu^+\nu_\mu$ and the control samples have been considered.  The
uncertainty of the $E_{\rm extra~\gamma}^{\rm max}$ requirement is
estimated to be 0.1\% by analyzing the DT hadronic events.  The
uncertainty in the $M_{K^-\mu^+}$ requirement is estimated with the
alternative $M_{K^-\mu^+}$ requirements of 1.51 or 1.61 GeV/$c^2$, and
the larger change on the BF 0.4\% is taken as the systematic
uncertainty.  The uncertainty of the $U_{\rm miss}$ fit is estimated
to be 0.5\% by applying different fit ranges, and signal and
background shapes.  The uncertainty of the limited MC size is 0.1\%.
The uncertainty in the MC model is estimated to be 0.1\%, which is the
difference between our nominal DT efficiency and that determined by
reweighting the $q^2$ distribution of the signal MC events to data
with the obtained FF parameters~(See below). The total uncertainty is
1.02\%, which is obtained by adding these uncertainties in quadrature.

The BFs of $D^0\to K^-\mu^+\nu_\mu$ and $\bar{D}^0\to
K^+\mu^-\bar{\nu}_\mu$ are measured separately. The results are
$\mathcal{B}_{D^0\to K^-\mu^+\nu_\mu}=(3.433\pm0.026_{\rm
  stat.}\pm0.039_{\rm syst.})\%$ and $\mathcal{B}_{\bar{D}^0\to
  K^+\mu^-\bar{\nu}_\mu}=(3.392\pm0.027_{\rm stat.}\pm0.034_{\rm
  syst.})\%$. 
The BF asymmetry is determined to be ${\mathcal
  A}=\frac{{\mathcal B}_{D^0\to K^-\mu^+\nu_\mu}-{\mathcal
    B}_{\bar{D}^0\to K^+\mu^-\bar{\nu}_\mu}}{{\mathcal B}_{D^0\to
    K^-\mu^+\nu_\mu}+{\mathcal B}_{\bar{D}^0\to
    K^+\mu^-\bar{\nu}_\mu}}=(0.6\pm0.6_{\rm stat.}\pm0.8_{\rm
  syst.})\%$, 
and no asymmetry in the BFs of $D^0\to K^-\mu^+\nu_\mu$ and $\bar{D}^0\to K^+\mu^-\bar\nu_\mu$ decays is found.  
All the systematic uncertainties except for those
in the $E_{\rm extra~\gamma}^{\rm max}$ requirement and MC model are studied
separately and are not canceled out in the BF asymmetry calculation.

\begin{figure*}[htbp]\centering
\includegraphics[width=0.9\textwidth]{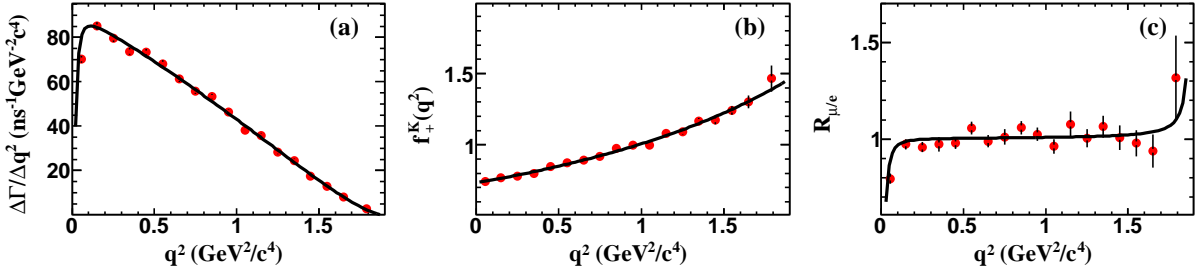}
\caption{(a) Fit to the PDRs, (b) projection to $f_+^K(q^2)$ for $D^0\to K^-\mu^+\nu_\mu$ and (c) the measured ${\mathcal R}_{\mu/e}$ in each $q^2$ interval. Dots with error bars are data. Solid curves are the fit, the projection or the ${\mathcal R}_{\mu/e}$ expected with the parameters in Ref.~\cite{ETM2017} where the uncertainty is negligible due to strong correlations in hadronic FFs.}
\label{fig:fitdecayrate}
\end{figure*}

The $D^0\to K^-\mu^+\nu_\mu$ dynamics is studied by dividing the SL
candidate events into various $q^2$ intervals.  The measured partial decay
rate~(PDR) in the $i$-th $q^2$ interval, $\Delta\Gamma_{\rm msr}^i$,
is determined by
\begin{equation}
	\Delta\Gamma^{i}_{\rm msr}\equiv\int_i(d\Gamma/dq^2)dq^2=N_{\mathrm{pro}}^{i}/(\tau_{D^{0}}\times N_{\mathrm{ST}}^{\rm tot}),
\end{equation}
where $N_{\mathrm{pro}}^{i}$ is the SL decay signal yield produced in
the $i$-th $q^{2}$ interval, $\tau_{D^{0}}$ is the $D^{0}$ lifetime
and $N_{\mathrm{ST}}^{\rm tot}$ is the ST yield.  The signal yield
produced in the $i$-th $q^{2}$ interval in data is calculated by
\begin{equation}
	N_{\mathrm{pro}}^{i}=\sum_{j}^{N_{\mathrm{intervals}}}(\varepsilon^{-1})_{ij}N_{\mathrm{obs}}^{j},
\end{equation}
where the observed DT yield in the $j$-th $q^{2}$ interval $N_{\rm
  obs}^j$ is obtained from the similar fit to the corresponding
$U_{\mathrm{miss}}$ distribution of data (See Fig.~2 of
Ref.~\cite{app}).  $\varepsilon$ is the efficiency matrix (Table~1 of
Ref.~\cite{app}), which is obtained by analyzing the signal MC events
and is given by
\begin{equation}
	\varepsilon_{ij}=\sum_k(1/N_{\rm ST}^{\rm tot})\times[(N^{ij}_{\mathrm{rec}}\times N_{\rm ST})/(N^{j}_{\mathrm{gen}}\times\varepsilon_{\mathrm{ST}})]_k,
\end{equation}
where $N^{ij}_{\mathrm{rec}}$ is the DT yield generated in the $j$-th
$q^{2}$ interval and reconstructed in the $i$-th $q^{2}$ interval,
$N^{j}_{\mathrm{gen}}$ is the total signal yield generated in the
$j$-th $q^{2}$ interval, and the index $k$ denotes the $k$-th ST mode.
The measured PDRs are shown in Fig.~\ref{fig:fitdecayrate}~(a) and
details can be found in Table~2 of Ref.~\cite{app}.

The FF is parametrized as the series expansion
parameterization~\cite{SEM} (SEP), which has been shown to be
consistent with constraints from
QCD~\cite{bes2015,cleo2009,babar2015}.
The 2-parameter SEP is chosen and is given by
\begin{equation}
	f^K_{+}(t)=\frac{1}{P(t)\Phi(t,t_{0})}\frac{f^K_{+}(0)P(0)\Phi(0,t_{0})}{1+r_{1}(t_{0})z(0,t_{0})}(1+r_{1}(t_{0})[z(t,t_{0})]).
\end{equation}
Here, $P(t)=z(t,m_{D^{*}_{s}}^{2})$ and $\Phi$ is given by
\begin{equation}
\begin{array}{l}
	\displaystyle \Phi(t,t_{0})=\sqrt{\frac{1}{24\pi\chi_{V}}}(\frac{t_{+}-t}{t_{+}-t_{0}})^{1/4}(\sqrt{t_{+}-t}+\sqrt{t_{+}})^{-5}\\
	\displaystyle \times(\sqrt{t_{+}-t}+\sqrt{t_{+}-t_{0}})(\sqrt{t_{+}-t}+\sqrt{t_{+}-t_{-}})^{3/2}\\
	\displaystyle \times(t_{+}-t)^{3/4},
\end{array}
\end{equation}
where
$z(t,t_{0})=\frac{\sqrt{t_{+}-t}-\sqrt{t_{+}-t_{0}}}{\sqrt{t_{+}-t}+\sqrt{t_{+}-t_{0}}}$,
$t_{\pm}=(m_{D}\pm m_{K})^{2}$, $t_{0}=t_{+}(1-\sqrt{1-t_{-}/t_{+}})$,
$m_D$ and $m_K$ are the masses of $D$ and $K$ particles,
$m_{D_s^*}$ is the pole mass of the vector FF accounting for the strong interaction between
$D$ and $K$ mesons and usually taken as the mass
of the lowest lying $c\bar s$ vector meson $D_s^*$~\cite{pdg2016},
 and $\chi_{V}$ can be obtained from dispersion
relations using perturbative QCD~\cite{chiV}.

The PDRs are fitted by assuming the ratio $f_{+}^K(q^{2})/f_{-}^K(q^{2})$ to be independent of $q^{2}$, and minimizing the $\chi^{2}$ constructed as
\begin{equation}
	\chi^{2}=\sum_{i,j=1}^{N_{\mathrm{intervals}}}(\Delta\Gamma^{i}_{\mathrm{msr}}-\Delta\Gamma^{i}_{\mathrm{exp}})
	C_{ij}^{-1}(\Delta\Gamma^{j}_{\mathrm{msr}}-\Delta\Gamma^{j}_{\mathrm{exp}}),
\end{equation}
where $\Delta\Gamma^i_{\rm exp}$ is the expected PDR in the $i$-th
$q^2$ interval given by~\cite{ddr,ddr2}
\begin{align}
	\Delta\Gamma^i_{\rm exp}&=\int_i\frac{G_{F}^{2}|V_{cs}|^{2}}{8\pi^{3}m_{D}}|\vec p_{K}||f_{+}^{K}(q^{2})|^{2}(\frac{W_{0}-E_{K}}{F_{0}})^{2}\nonumber\\
	&\times\lbrack\frac{1}{3}m_{D}|\vec p_{K}|^{2}+\frac{m_{\ell}^{2}}{8m_{D}}(m_{D}^{2}+m_{K}^{2}+2m_{D}E_{K})\nonumber\\
	&+\frac{1}{3}m_{\ell}^{2}\frac{|\vec p_{K}|^{2}}{F_{0}}+\frac{1}{4}m_{\ell}^{2}\frac{m_{D}^{2}-m_{K}^{2}}{m_{D}}\mathrm{Re}(\frac{f^K_{-}(q^{2})}{f^K_{+}(q^{2})})\nonumber\\
	&+\frac{1}{4}m_{\ell}^{2}F_{0}|\frac{f^K_{-}(q^{2})}{f^K_{+}(q^{2})}|^{2}\rbrack dq^2,
\label{eq:ddr}
\end{align}
and $C_{ij} = C_{ij}^{\mathrm{stat}}+C_{ij}^{\mathrm{syst}}$ is the covariance
matrix of the measured PDRs among $q^2$ intervals. 
In Eq.~(\ref{eq:ddr}), $G_F$ is the Fermi coupling constant; $m_\ell$ is
the mass of the lepton;
$|\vec p_{K}|$ and $E_{K}$ are the
momentum and energy of the kaon in the $D$ rest frame;
$W_{0}=(m_{D}^{2}+m_{K}^{2}-m_{\ell}^{2})/(2m_{D})$ is the maximum energy of the kaon in the $D$ rest frame; and
$F_{0}=W_{0}-E_{K}+m_{\ell}^{2}/(2m_{D})=q^2/(2m_D)$.  
The statistical covariance matrix~(Table 3 of Ref.~\cite{app}) is constructed as
\begin{equation}
	C_{ij}^{\rm stat} = (\frac{1}{\tau_{D^{0}}N_{\mathrm{ST}}^{\rm tot}})^{2}\sum_{\alpha}\varepsilon_{i\alpha}^{-1}\varepsilon_{j\alpha}^{-1}(\sigma(N_{\mathrm{obs}}^{\alpha}))^{2}.
\end{equation}
The systematic covariance matrix (Table~4 of Ref.~\cite{app}) is obtained by summing all the covariance matrices for each source of systematic uncertainty. In general, it has the form
\begin{equation}
C_{ij}^{\mathrm{syst}}=\delta(\Delta\Gamma^{i}_{\rm msr})\delta(\Delta\Gamma^{j}_{\rm msr}),
\end{equation}
where $\delta(\Delta\Gamma^{i}_{\rm msr})$ is the systematic
uncertainty of the PDR in the $i$-th $q^{2}$ interval. The systematic
uncertainties in $N_{\rm ST}^{\rm tot}$, $\tau_{D^0}$ and $E_{\rm
  extra~\gamma}^{\rm max}$ requirement are considered to be fully
correlated across $q^2$ intervals while others are studied separately
in each $q^2$ interval with the same method used in the BF
measurement.

Figures~\ref{fig:fitdecayrate}(a) and (b) show the fit to the PDRs of
$D^0\to K^-\mu^+\nu_\mu$ and the projection to $f^K_+(q^2)$. The
goodness of fit is $\chi^2/{\rm NDOF}=15.0/15$, where NDOF is the
number of degrees of freedom.  From the fit, we obtain the product of
$f_+^K(0)|V_{cs}|=0.7133\pm0.0038_{\rm stat.}\pm0.0030_{\rm syst.}$,
the first order coefficient $r_1=-1.90\pm0.21_{\rm stat.}\pm0.07_{\rm
  syst.}$ and the FF ratio $f_-^K/f_+^K=-0.6\pm0.8_{\rm
  stat.}\pm0.2_{\rm syst.}$.  The nominal fit parameters are taken
from the results obtained by fitting with the combined statistical and
systematic covariance matrix, and the statistical uncertainties of the
fit parameters are taken from the fit with only the statistical covariance
matrix. For each parameter, the systematic uncertainty is obtained by
calculating the quadratic difference of uncertainties between these
two fits.

Combining ${\mathcal B}_{D^0\to K^-\mu^+\nu_\mu}$ with our previous
measurement ${\mathcal B}_{D^0\to K^-e^+\nu_e}=(3.505\pm0.014_{\rm
  stat.}\pm0.033_{\rm syst.})\%$~\cite{bes2015} gives
${\mathcal R}_{\mu/e}=0.974\pm0.007_{\rm stat.}\pm0.012_{\rm syst.}$, which
agrees with the theoretical calculations with LQCD~\cite{ETM2017,Riggio2018} and an SM quark
model~\cite{Soni2017}.  Additionally, we determine ${\mathcal R}_{\mu/e}$ in each
$q^2$ interval, as shown in Fig.~\ref{fig:fitdecayrate}(c), where the
error bars include both statistical and the uncanceled systematic
uncertainties. In the ${\mathcal R}_{\mu/e}$ calculation, the uncertainties in
$N_{\rm ST}^{\rm tot}$, $\tau_{D^0}$ as well as the tracking and PID
efficiencies of the kaon cancel.
Below $q^2=0.1$ GeV$^2/c^4$, ${\mathcal R}_{\mu/e}$ is significantly lower than 1 
due to smaller phase space for $D^0\to K^-\mu^+\nu_\mu$
with nonzero muon mass that cannot be neglected. Above $0.1$ GeV$^2/c^4$,
${\mathcal R}_{\mu/e}$ is close to 1. They are consistent with the SM prediction, 
and no deviation larger than 2$\sigma$ is observed.

In summary, by analyzing $2.93~\mathrm{fb}^{-1}$ of data collected at
$\sqrt{s}=3.773$ GeV with the BESIII detector,
we present an improved measurement of the absolute BF of the SL decay
$D^{0}\to K^-\mu^+\nu_\mu$. Our result is consistent with the PDG
value~\cite{pdg2016} and improves its precision by a factor of three.
Combining the previous BESIII measurements of $D^0\to K^-e^+\nu_e$, we
calculate ${\mathcal R}_{\mu/e}$ ratios in the full $q^2$ range and various $q^2$ intervals. No
significant evidence of LFU violation is found with current statistics and systematic uncertainties.
By fitting the PDRs of this decay, we obtain
$f_{+}^{K}(0)|V_{cs}|=0.7133\pm0.0038_{\rm stat.}\pm0.0029_{\rm
  syst.}$. Using $|V_{cs}|$ given by global fit in the SM~\cite{pdg2016} yields
$f_{+}^{K}(0)=0.7327\pm0.0039_{\rm stat.}\pm0.0030_{\rm syst.}$, while
using the $f_{+}^{K}(0)$ calculated in LQCD~\cite{LQCD} results in
$|V_{cs}| = 0.955\pm0.005_{\rm stat.}\pm0.004_{\rm
  syst.}\pm0.024_{\rm LQCD}$. These results are consistent with our
measurements using $D^{0(+)}\to\bar Ke^+\nu_e$~\cite{bes2015,bes3_ksev,bes3_klev} and $D_s^+\to\mu^+\nu_\mu$~\cite{bes3_muv} within
uncertainties and are important to test the LQCD calculation of
$f^K_+(0)$~\cite{LQCD,MILC,ETM2017} and quark mixing matrix unitarity with better accuracy.

The BESIII collaboration thanks the staff of BEPCII and the IHEP computing center for their strong support. This work is supported in part by National Key Basic Research Program of China under Contract No. 2015CB856700;
National Natural Science Foundation of China (NSFC) under Contracts Nos. 11305180, 11775230, 11235011, 11335008, 11425524, 11625523, 11635010; 
the Chinese Academy of Sciences (CAS) Large-Scale Scientific Facility Program; 
the CAS Center for Excellence in Particle Physics (CCEPP); 
Joint Large-Scale Scientific Facility Funds of the NSFC and CAS under Contracts Nos. U1632109, U1332201, U1532257, U1532258; 
CAS under Contracts Nos. KJCX2-YW-N29, KJCX2-YW-N45, QYZDJ-SSW-SLH003; 
100 Talents Program of CAS; 
National 1000 Talents Program of China; 
INPAC and Shanghai Key Laboratory for Particle Physics and Cosmology; 
German Research Foundation DFG under Contracts Nos. Collaborative Research Center CRC 1044, FOR 2359; 
Istituto Nazionale di Fisica Nucleare, Italy; 
Koninklijke Nederlandse Akademie van Wetenschappen (KNAW) under Contract No. 530-4CDP03; 
Ministry of Development of Turkey under Contract No. DPT2006K-120470; 
National Science and Technology fund; 
The Swedish Research Council; 
U. S. Department of Energy under Contracts Nos. DE-FG02-05ER41374, DE-SC-0010118, DE-SC-0010504, DE-SC-0012069; 
University of Groningen (RuG) and the Helmholtzzentrum fuer Schwerionenforschung GmbH (GSI), Darmstadt; 
WCU Program of National Research Foundation of Korea under Contract No. R32-2008-000-10155-0.

\clearpage
\appendix
\onecolumngrid
\section*{Supplemental material}
\setcounter{table}{0}
\setcounter{figure}{0}

\begin{figure*}[htbp]\centering
\includegraphics[width=0.6\textwidth]{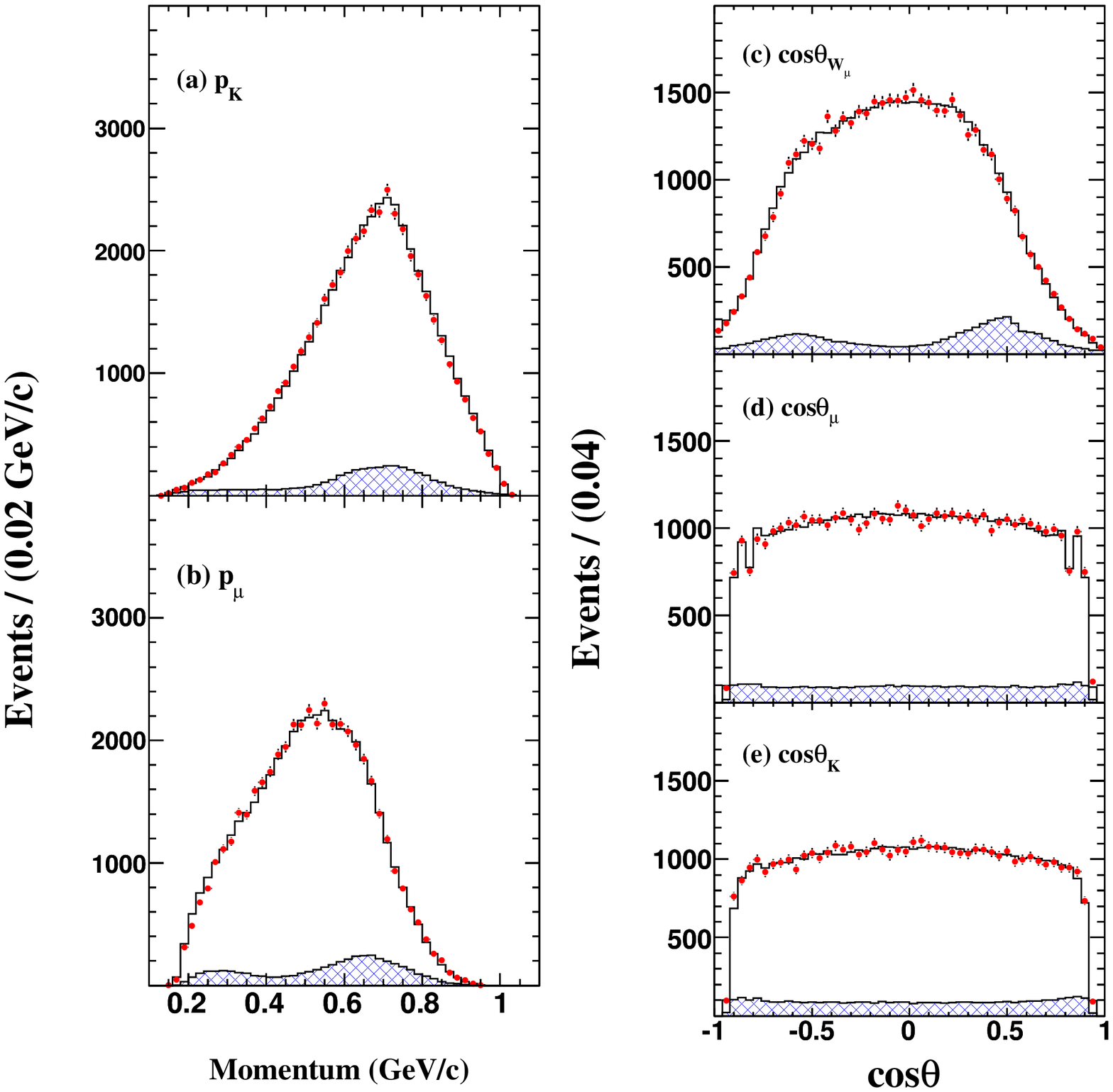}
\caption{Comparisons between data and MC simulation of distributions of the kaon and muon momentum and $\cos\theta$ as well as $\cos\theta_{W_\mu}$ for $D^0\to K^-\mu^+\nu_\mu$ candidate events, where $\theta_{W_\mu}$ is the angle between the direction of the virtual $W^+$ boson in the $D^0$ rest frame and the momentum of muon in the $W^+$ rest frame. These events satisfy $-0.06<U_{\rm miss}<0.02$~GeV. The red dots with error bars denote data, the solid histograms are the MC simulated signal plus background and the cross-hatched histograms are the MC simulated background only.}
\label{fig:comparison}
\end{figure*}

Figure~\ref{fig:comparison} shows the comparisons of some typical distributions for $D^0\to K^-\mu^+\nu_\mu$ candidate events between data and MC simulation.

\begin{figure*}[htbp]\centering
\includegraphics[width=\textwidth]{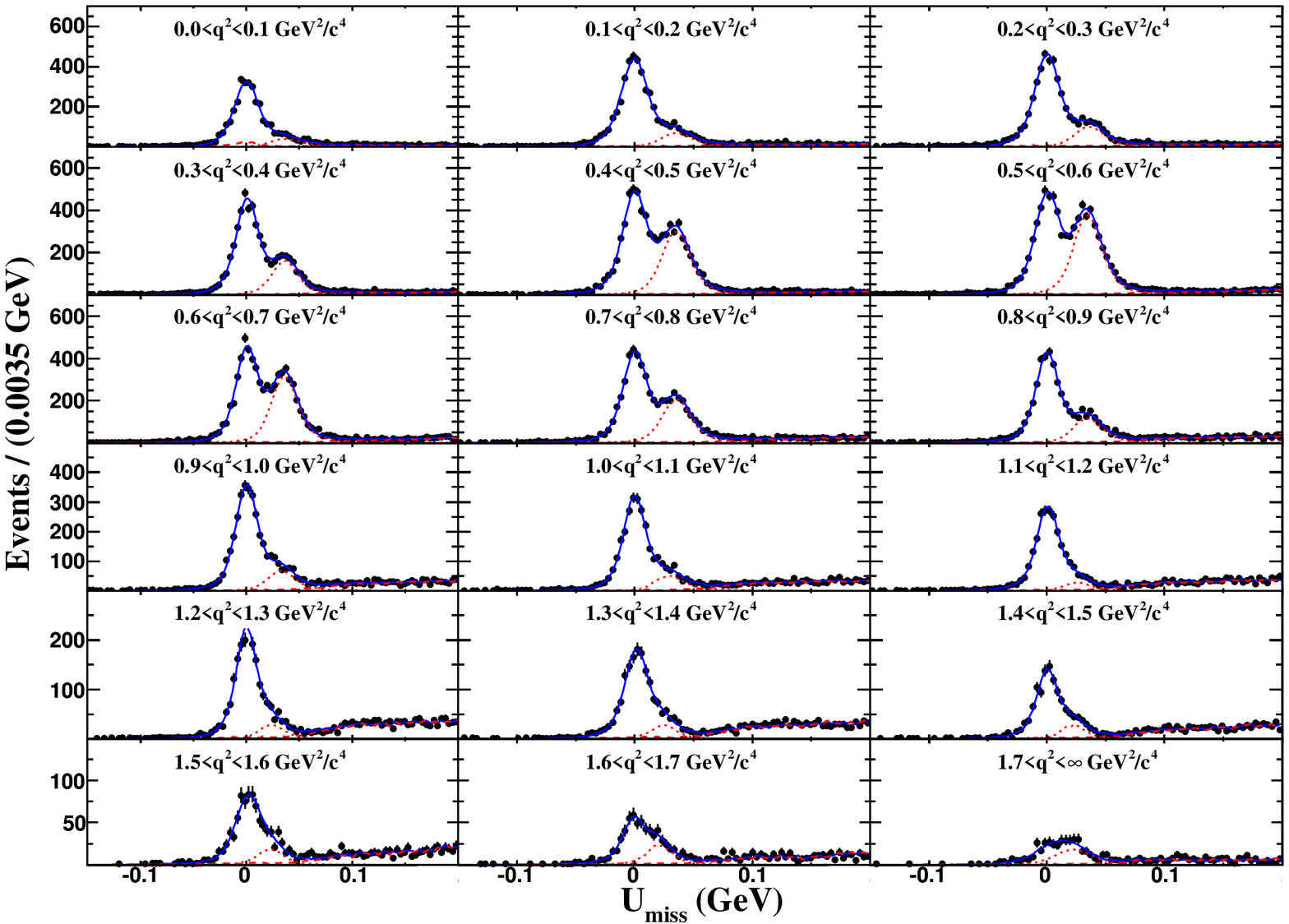}
\caption{Fits to $U_{\rm miss}$ distributions in each reconstructed $q^2$ bins of data, where the dots with error bars are data, the blue solid curve is the best fit, the red dotted curve is the $D^0\to K^-\pi^+\pi^0$ peaking background and the red dashed curve is the combinatorial background.}
\label{fig:umissfit}
\end{figure*}

Figure~\ref{fig:umissfit} shows the fits to the $U_{\rm miss}$ distributions for $D^0\to K^-\mu^+\nu_\mu$ candidate events of data in 18 $q^2$ intervals.

Table~\ref{tab:effmatrix} gives the weighted efficiency matrix for all three single tag modes for the reconstruction of $D^0\to K^-\mu^+\nu_\mu$ events.

Table~\ref{tab:decayrate} presents the number of reconstructed events $N_{\rm obs}^i$ obtained from the $U_{\rm miss}$ fits as shown in Fig.~\ref{fig:umissfit}, the number of produced events $N_{\rm pro}^i$, the measured PDR $\Delta\Gamma_{\rm msr}^i$ and ${\mathcal R}_{\mu/e}$ in each $q^2$ interval.

Tables~\ref{tab:statcov} and \ref{tab:systcov} summarize the statistical and systematic covariance matrices for the measured PDRs in different $q^2$ intervals, respectively.

\begin{table*}[htbp]\centering
\caption{Weighted efficiency matrix for all three single tag modes, where $\varepsilon_{ij}$ represents the 
efficiency of events generated in the $j$-th $q^2$ interval and reconstructed in the $i$-th $q^2$ interval.}
\label{tab:effmatrix}
\begin{tabular}{ccccccccccccccccccc}\hline\hline
$\varepsilon_{ij}~(\%)$&1&2&3&4&5&6&7&8&9&10&11&12&13&14&15&16&17&18\\
\hline
1&45.49&1.35&0.02&0.01&0.00&0.00&0.00&0.00&0.00&0.00&0.00&0.00&0.00&0.00&0.00&0.00&0.00&0.00\\
2&1.80&45.09&2.05&0.03&0.01&0.00&0.00&0.00&0.00&0.00&0.00&0.00&0.00&0.00&0.00&0.00&0.00&0.00\\
3&0.04&1.95&46.76&2.56&0.05&0.01&0.01&0.00&0.00&0.00&0.00&0.00&0.00&0.00&0.00&0.00&0.00&0.00\\
4&0.02&0.06&2.48&49.30&3.01&0.08&0.02&0.01&0.00&0.00&0.00&0.00&0.00&0.00&0.00&0.00&0.00&0.00\\
5&0.01&0.02&0.09&2.95&51.96&3.31&0.10&0.02&0.01&0.01&0.00&0.00&0.00&0.00&0.00&0.00&0.00&0.00\\
6&0.00&0.01&0.03&0.11&3.33&54.37&3.57&0.12&0.04&0.02&0.01&0.00&0.00&0.00&0.00&0.00&0.00&0.00\\
7&0.00&0.01&0.02&0.04&0.13&3.66&56.65&3.80&0.14&0.04&0.02&0.01&0.00&0.00&0.00&0.00&0.00&0.00\\
8&0.00&0.00&0.01&0.02&0.05&0.17&3.92&58.23&3.78&0.17&0.06&0.03&0.01&0.00&0.00&0.00&0.00&0.00\\
9&0.00&0.00&0.01&0.01&0.02&0.06&0.19&4.07&59.44&3.89&0.17&0.07&0.02&0.01&0.00&0.00&0.00&0.00\\
10&0.00&0.00&0.00&0.01&0.01&0.02&0.07&0.20&4.04&59.52&3.72&0.19&0.07&0.03&0.00&0.00&0.00&0.00\\
11&0.00&0.00&0.00&0.00&0.01&0.02&0.03&0.07&0.22&3.96&59.13&3.61&0.19&0.06&0.01&0.01&0.00&0.00\\
12&0.00&0.00&0.00&0.00&0.00&0.01&0.01&0.03&0.08&0.24&3.87&58.83&3.36&0.20&0.06&0.01&0.00&0.00\\
13&0.00&0.00&0.00&0.00&0.00&0.00&0.01&0.01&0.03&0.07&0.25&3.73&57.92&3.16&0.16&0.04&0.00&0.00\\
14&0.00&0.00&0.00&0.00&0.00&0.00&0.00&0.00&0.01&0.02&0.07&0.24&3.48&56.60&2.94&0.12&0.02&0.00\\
15&0.00&0.00&0.00&0.00&0.00&0.00&0.00&0.00&0.00&0.00&0.02&0.06&0.24&3.35&55.35&2.59&0.09&0.00\\
16&0.00&0.00&0.00&0.00&0.00&0.00&0.00&0.00&0.00&0.00&0.01&0.01&0.05&0.19&3.01&52.79&2.25&0.06\\
17&0.00&0.00&0.00&0.00&0.00&0.00&0.00&0.00&0.00&0.00&0.00&0.00&0.01&0.03&0.14&2.47&49.49&1.63\\
18&0.00&0.00&0.00&0.00&0.00&0.00&0.00&0.00&0.00&0.00&0.00&0.00&0.00&0.00&0.01&0.07&1.80&36.80\\
\hline
\end{tabular}
\end{table*}

\begin{table*}[htbp]\centering
\caption{The PDR of $D^0\to K^-\mu^+\nu_\mu$ and ${\mathcal R}_{\mu/e}$ in each $q^{2}$ bin of data, where uncertainties of PRDs are statistical only.}
\label{tab:decayrate}
\begin{tabular}{cccccc}\hline
$i$ & \parbox{1.5cm}{$q^{2}$ $(\mathrm{GeV}^{2}/c^{4})$}&$N_{\mathrm{obs}}^i$&$N_{\mathrm{pro}}^i$&\parbox{1cm}{$\Delta\Gamma^i_{\rm msr}$ $(\mathrm{ns^{-1}})$} & ${\mathcal R}_{\mu/e}$\\
\hline
1  & (0.0,\,0.1)     &2834.1$\pm$63.3&5984.4$\pm$139.3&6.232$\pm$0.145 & 0.796$\pm$0.027\\
2  & (0.1,\,0.2)     &3952.1$\pm$71.0&8172.3$\pm$158.1&8.511$\pm$0.165 & 0.973$\pm$0.026\\
3  & (0.2,\,0.3)     &3918.6$\pm$70.7&7636.2$\pm$152.2&7.953$\pm$0.158 & 0.959$\pm$0.026\\
4  & (0.3,\,0.4)     &3901.2$\pm$71.8&7073.8$\pm$147.0&7.367$\pm$0.153 & 0.974$\pm$0.037\\
5  & (0.4,\,0.5)     &4099.6$\pm$77.6&7037.5$\pm$150.9&7.329$\pm$0.157 & 0.979$\pm$0.029\\
6  & (0.5,\,0.6)     &4024.5$\pm$78.4&6545.0$\pm$145.9&6.816$\pm$0.152 & 1.057$\pm$0.034\\
7  & (0.6,\,0.7)     &3806.2$\pm$75.2&5892.6$\pm$134.5&6.137$\pm$0.140 & 0.990$\pm$0.031\\
8  & (0.7,\,0.8)     &3575.2$\pm$70.2&5363.0$\pm$122.3&5.585$\pm$0.127 & 1.012$\pm$0.039\\
9  & (0.8,\,0.9)     &3460.2$\pm$67.4&5115.8$\pm$114.9&5.328$\pm$0.120 & 1.060$\pm$0.034\\
10 & (0.9,\,1.0)     &3022.3$\pm$64.1&4455.8$\pm$109.1&4.640$\pm$0.114 & 1.026$\pm$0.035\\
11 & (1.0,\,1.1)     &2497.2$\pm$58.4&3671.0$\pm$100.1&3.823$\pm$0.104 & 0.963$\pm$0.036\\
12 & (1.1,\,1.2)     &2279.4$\pm$60.4&3437.4$\pm$103.9&3.580$\pm$0.108 & 1.076$\pm$0.068\\
13 & (1.2,\,1.3)     &1801.0$\pm$54.6&2727.6$\pm$95.3 &2.841$\pm$0.099 & 1.004$\pm$0.047\\
14 & (1.3,\,1.4)     &1483.7$\pm$52.0&2340.3$\pm$92.8 &2.437$\pm$0.097 & 1.065$\pm$0.057\\
15 & (1.4,\,1.5)     &1051.2$\pm$45.6&1680.4$\pm$83.1 &1.750$\pm$0.087 & 1.008$\pm$0.064\\
16 & (1.5,\,1.6)     &727.1$\pm$32.1 &1235.5$\pm$61.4 &1.287$\pm$0.064 & 0.979$\pm$0.067\\
17 & (1.6,\,1.7)     &425.2$\pm$25.9 &774.3$\pm$52.7  &0.806$\pm$0.055 & 0.940$\pm$0.086\\
18 & (1.7,\,$\infty$)&191.7$\pm$22.0 &479.5$\pm$59.9  &0.499$\pm$0.062 & 1.318$\pm$0.217\\
\hline
\end{tabular}
\end{table*}

\begin{table*}[htbp]\centering
\caption{Statistical covariance density matrix for the measured PDRs of $D^0\to K^-\mu^+\nu_\mu$ in different
$q^2$ intervals.}
\label{tab:statcov}
\begin{tabular}{ccccccccccccccccccc}\hline\hline
$\rho_{ij}$&1&2&3&4&5&6&7&8&9&10&11&12&13&14&15&16&17&18\\
\hline
1&1.000&-0.069&0.003&-0.001&0.000&0.000&0.000&0.000&0.000&0.000&0.000&0.000&0.000&0.000&0.000&0.000&0.000&0.000\\
2&-0.069&1.000&-0.087&0.005&-0.001&0.000&0.000&0.000&0.000&0.000&0.000&0.000&0.000&0.000&0.000&0.000&0.000&0.000\\
3&0.003&-0.087&1.000&-0.105&0.006&-0.001&0.000&0.000&0.000&0.000&0.000&0.000&0.000&0.000&0.000&0.000&0.000&0.000\\
4&-0.001&0.005&-0.105&1.000&-0.117&0.008&-0.001&0.000&0.000&0.000&0.000&0.000&0.000&0.000&0.000&0.000&0.000&0.000\\
5&0.000&-0.001&0.006&-0.117&1.000&-0.124&0.008&-0.002&0.000&0.000&0.000&0.000&0.000&0.000&0.000&0.000&0.000&0.000\\
6&0.000&0.000&-0.001&0.008&-0.124&1.000&-0.130&0.008&-0.002&0.000&0.000&0.000&0.000&0.000&0.000&0.000&0.000&0.000\\
7&0.000&0.000&0.000&-0.001&0.008&-0.130&1.000&-0.134&0.008&-0.002&0.000&0.000&0.000&0.000&0.000&0.000&0.000&0.000\\
8&0.000&0.000&0.000&0.000&-0.002&0.008&-0.134&1.000&-0.133&0.007&-0.002&0.000&0.000&0.000&0.000&0.000&0.000&0.000\\
9&0.000&0.000&0.000&0.000&0.000&-0.002&0.008&-0.133&1.000&-0.132&0.007&-0.002&0.000&0.000&0.000&0.000&0.000&0.000\\
10&0.000&0.000&0.000&0.000&0.000&0.000&-0.002&0.007&-0.132&1.000&-0.129&0.005&-0.002&0.000&0.000&0.000&0.000&0.000\\
11&0.000&0.000&0.000&0.000&0.000&0.000&0.000&-0.002&0.007&-0.129&1.000&-0.125&0.005&-0.002&0.000&0.000&0.000&0.000\\
12&0.000&0.000&0.000&0.000&0.000&0.000&0.000&0.000&-0.002&0.005&-0.125&1.000&-0.121&0.003&-0.001&0.000&0.000&0.000\\
13&0.000&0.000&0.000&0.000&0.000&0.000&0.000&0.000&0.000&-0.002&0.005&-0.121&1.000&-0.115&0.003&-0.001&0.000&0.000\\
14&0.000&0.000&0.000&0.000&0.000&0.000&0.000&0.000&0.000&0.000&-0.002&0.003&-0.115&1.000&-0.113&0.004&-0.001&0.000\\
15&0.000&0.000&0.000&0.000&0.000&0.000&0.000&0.000&0.000&0.000&0.000&-0.001&0.003&-0.113&1.000&-0.111&0.002&0.000\\
16&0.000&0.000&0.000&0.000&0.000&0.000&0.000&0.000&0.000&0.000&0.000&0.000&-0.001&0.004&-0.111&1.000&-0.094&0.002\\
17&0.000&0.000&0.000&0.000&0.000&0.000&0.000&0.000&0.000&0.000&0.000&0.000&0.000&-0.001&0.002&-0.094&1.000&-0.080\\
18&0.000&0.000&0.000&0.000&0.000&0.000&0.000&0.000&0.000&0.000&0.000&0.000&0.000&0.000&0.000&0.002&-0.080&1.000\\
\hline
\end{tabular}
\end{table*}

\begin{table*}[htbp]\centering
\caption{Systematic covariance density matrix for the measured PDRs of $D^0\to K^-\mu^+\nu_\mu$ in different $q^2$ intervals.}
\label{tab:systcov}
\begin{tabular}{ccccccccccccccccccc}\hline\hline
$\rho_{ij}$&1&2&3&4&5&6&7&8&9&10&11&12&13&14&15&16&17&18\\
\hline
1&1.000&0.151&0.135&-0.475&0.729&-0.265&0.776&-0.454&0.264&0.435&0.394&0.085&0.326&0.283&0.300&0.256&0.270&0.186\\
2&0.151&1.000&0.791&0.572&0.469&0.720&0.345&0.591&0.778&0.650&0.612&0.133&0.508&0.441&0.468&0.399&0.420&0.287\\
3&0.135&0.791&1.000&0.635&0.450&0.784&0.314&0.661&0.811&0.663&0.625&0.135&0.519&0.451&0.478&0.408&0.430&0.294\\
4&-0.475&0.572&0.635&1.000&-0.229&0.925&-0.383&0.975&0.497&0.219&0.220&0.048&0.185&0.161&0.171&0.146&0.154&0.106\\
5&0.729&0.469&0.450&-0.229&1.000&0.026&0.877&-0.194&0.563&0.675&0.620&0.134&0.514&0.447&0.474&0.406&0.429&0.296\\
6&-0.265&0.720&0.784&0.925&0.026&1.000&-0.136&0.933&0.673&0.424&0.409&0.089&0.342&0.297&0.316&0.271&0.286&0.198\\
7&0.776&0.345&0.314&-0.383&0.877&-0.136&1.000&-0.362&0.451&0.604&0.553&0.119&0.458&0.398&0.424&0.363&0.385&0.268\\
8&-0.454&0.591&0.661&0.975&-0.194&0.933&-0.362&1.000&0.512&0.246&0.244&0.053&0.205&0.179&0.190&0.164&0.174&0.121\\
9&0.264&0.778&0.811&0.497&0.563&0.673&0.451&0.512&1.000&0.675&0.667&0.141&0.555&0.482&0.514&0.442&0.470&0.329\\
10&0.435&0.650&0.663&0.219&0.675&0.424&0.604&0.246&0.675&1.000&0.598&0.143&0.533&0.465&0.496&0.427&0.455&0.319\\
11&0.394&0.612&0.625&0.220&0.620&0.409&0.553&0.244&0.667&0.598&1.000&-0.135&0.578&0.435&0.465&0.400&0.427&0.299\\
12&0.085&0.133&0.135&0.048&0.134&0.089&0.119&0.053&0.141&0.143&-0.135&1.000&-0.192&0.094&0.096&0.087&0.093&0.066\\
13&0.326&0.508&0.519&0.185&0.514&0.342&0.458&0.205&0.555&0.533&0.578&-0.192&1.000&0.299&0.395&0.338&0.363&0.257\\
14&0.283&0.441&0.451&0.161&0.447&0.297&0.398&0.179&0.482&0.465&0.435&0.094&0.299&1.000&0.262&0.299&0.320&0.228\\
15&0.300&0.468&0.478&0.171&0.474&0.316&0.424&0.190&0.514&0.496&0.465&0.096&0.395&0.262&1.000&0.251&0.350&0.249\\
16&0.256&0.399&0.408&0.146&0.406&0.271&0.363&0.164&0.442&0.427&0.400&0.087&0.338&0.299&0.251&1.000&0.234&0.224\\
17&0.270&0.420&0.430&0.154&0.429&0.286&0.385&0.174&0.470&0.455&0.427&0.093&0.363&0.320&0.350&0.234&1.000&0.192\\
18&0.186&0.287&0.294&0.106&0.296&0.198&0.268&0.121&0.329&0.319&0.299&0.066&0.257&0.228&0.249&0.224&0.192&1.000\\
\hline
\hline
\end{tabular}
\end{table*}


\begin{thebibliography}{99}

\bibitem{babar_1} J. P. Lees {\it et al.} (BaBar Collaboration), Phys. Rev. Lett.
{\bf 109}, 101802 (2012).
\bibitem{babar_2} J. P. Lees {\it et al.} (BaBar Collaboration), Phys. Rev. D {\bf 88},
072012 (2013).
\bibitem{lhcb_1} R. Aaij {\it et al.} (LHCb Collaboration), Phys. Rev. Lett. {\bf 115},
111803 (2015).
\bibitem{belle2015} M. Huschle {\it et al.} (Belle Collaboration), Phys. Rev. D {\bf 92},
072014 (2015).
\bibitem{belle2016} Y. Sato {\it et al.} (Belle Collaboration), Phys. Rev. D {\bf 94},
072007 (2016).

\bibitem{lhcb_kee_3}
R. Aaij {\it et al.} (LHCb Collaboration), Phys. Rev. Lett. {\bf 113}, 151601 (2014).
\bibitem{lhcb_kee_4}
R. Aaij {\it et al.} (LHCb Collaboration), JHEP {\bf 08}, 055 (2017).
\bibitem{HFLAV} Y. Amhis {\it et al.} (HFLAV Collaboration), Eur. Phys. J. C {\bf 77}, 895 (2017).

\bibitem{BFajfer2012} S. Fajfer, J. F. Kamenik and I. Nisandzic,
Phys. Rev. D. {\bf 85}, 094025 (2012).
\bibitem{Fajfer2012} S. Fajfer {\it et al.},
Phys. Rev. Lett. {\bf 109}, 161801 (2012).
\bibitem{Celis2013} A. Celis {\it et al.},
J. High Energy Phys. {\bf 1301}, 054 (2013).
\bibitem{Crivellin2015} A. Crivellin, G. D'Ambrosio and J. Heeck,
Phys. Rev. Lett. {\bf 114}, 151801 (2015).
\bibitem{Crivellin2016} A. Crivellin, J. Heeck and P. Stoffer,
Phys. Rev. Lett. {\bf 116}, 081801 (2016).
\bibitem{Bauer2016} M. Bauer and M. Neubert,
Phys. Rev. Lett. {\bf 116}, 141802 (2016).
\bibitem{Fajfer2015} S. Fajfer, I. Nisandzic and U. Rojec,
Phys. Rev. D {\bf 91}, 094009 (2015).
\bibitem{Riggio2018} L. Riggio, G. Salerno and S. Simula, Eur. Phys. J. C{\bf 78}, 501 (2018).
\bibitem{ETM2017} V. Lubicz {\it et al.} (ETM Collaboration), Phys. Rev. D {\bf 96}, 054514 (2017).
\bibitem{charge} Throughout this Letter, the charge conjugate channels are implied unless otherwise stated.
\bibitem{Zhang2018} J. Zhang, C. X. Yue and C. H. Li, Eur. Phys. J. C {\bf 78}, 695 (2018).
\bibitem{Fang2015} Y. Fang {\it et al.}, Eur. Phys. J. C {\bf 75}, 10 (2015).
\bibitem{belle2006} L. Widhalm {\it et al.} (Belle Collaboration), Phys. Rev. Lett. {\bf 97}, 061804 (2006).
\bibitem{cleo2009} D. Besson {\it et al.} (CLEO Collaboration), Phys. Rev. D {\bf 80}, 032005 (2009).
\bibitem{babar2007} B. Aubert {\it et al.} (BaBar Collaboration), Phys. Rev. D {\bf 76}, 052005 (2007).
\bibitem{bes2015} M. Ablikim {\it et al.} (BESIII Collaboration), Phys. Rev. D {\bf 92}, 072012 (2015).
\bibitem{focus2005} J. M. Link {\it et al.} (FOCUS Collaboration), Phys. Lett. B {\bf 607}, 233 (2005).
\bibitem{pdg2016} M. Tanabashi {\it et al.} (Particle Data Group), Phys. Rev. D {\bf 98}, 030001 (2018).
\bibitem{LQCD} H. Na {\it et al.} (HPQCD Collaboration), Phys. Rev. D {\bf 82}, 114506 (2010).
\bibitem{BESCol} M. Ablikim {\it et al.} (BESIII Collaboration), Nucl. Instr. Meth. A {\bf 614}, 345 (2010).
\bibitem{geant4} S. Agostinelli {\it et al.} ({\sc geant4} Collaboration), Nucl. Instr. Meth. A {\bf 506}, 250 (2003).
\bibitem{kkmc} S. Jadach, B. F. L. Ward and Z. Was, Comp. Phys. Commu. {\bf 130}, 260 (2000);
Phys. Rev. D {\bf 63}, 113009 (2001).
\bibitem{evtgen} D. J. Lange, Nucl. Instr. Meth. A {\bf 462}, 152 (2001);
R. G. Ping, Chin. Phys. C {\bf 32}, 599 (2008).
\bibitem{lundcharm}
J. C. Chen {\it et al.}, Phys. Rev. D {\bf 62}, 034003 (2000).
\bibitem{MPM} D. Becirevic and A. B. Kaidalov, Phys. Lett. B {\bf 478}, 417 (2000).
\bibitem{cosmic} M. Ablikim {\it et al.} (BESIII Collaboration), Phys. Lett. B {\bf 734}, 227 (2014).
\bibitem{argus} H. Albrecht {\it et al.} (ARGUS Collaboration), Phys. Lett. B {\bf 241}, 278 (1990).
\bibitem{app} See Supplemental Material at [URL will be inserted by publisher] for the comparisons of some typical distributions between data and MC simulation, efficiency matrix, fits to $U_{\rm miss}$ distributions in 18 $q^2$ intervals, PDR and ${\mathcal R}_{\mu/e}$ in each $q^2$ interval, statistical and systematic covariance matrices.
\bibitem{SEM} T. Becher and R. J. Hill, Phys. Lett. B {\bf 633}, 61 (2006).
\bibitem{babar2015} J. P. Lees {\it et al.} (BaBar Collaboration), Phys. Rev. D {\bf 91}, 052022 (2015).
\bibitem{chiV} C. G. Boyd, B. Grinstein and R. F. Lebed, Nucl. Phys. B {\bf 461}, 493 (1996).
\bibitem{ddr} J. M. Link {\it et al.} (FOCUS Collaboration), Phys. Lett. B {\bf 607}, 233 (2005).
\bibitem{ddr2} J. G. Korner and G. A. Schuler, Z. Phys. C {\bf 46}, 93 (1990).

\bibitem{Soni2017} N. R. Soni and J. N. Pandya,
Phys. Rev. D {\bf 96}, 016017 (2017).
\bibitem{bes3_ksev} M. Ablikim {\it et al.} (BESIII Collaboration), Phys. Rev. D {\bf 96}, 012002 (2017).
\bibitem{bes3_klev} M. Ablikim {\it et al.} (BESIII Collaboration), Phys. Rev. D {\bf 92}, 112008 (2015).
\bibitem{bes3_muv} M. Ablikim {\it et al.} (BESIII Collaboration), arXiv:1811.10890.
\bibitem{MILC} C. Aubin {\it et al.} (Fermilab Lattice and MILC and HPQCD Collaborations),
Phys. Rev. Lett. {\bf 94}, 011601 (2005).

\end{thebibliography}
\end{document}